\documentclass{mem}
\usepackage{natbib}\usepackage{txfonts}\usepackage{balance}
\usepackage{graphicx}
\usepackage[a4paper,breaklinks,dvipdfm]{hyperref}
\idline{75}{282}
\begin{document}
\def\teff{$T\rm_{eff }$}
\def\kms{$\mathrm {km s}^{-1}$}

\title{
Young Brown Dwarfs as Giant Exoplanet Analogs
}

   \subtitle{}

\author{
Jacqueline K. Faherty\inst{1,2}, Kelle L. Cruz\inst{2,4}, Emily L. Rice\inst{2,3}, Adric Riedel\inst{2,4}         }

  \offprints{J. K. Faherty}
 
\institute{
Department of Astronomy, 
Universidad de Chile Cerro Calan, Las Condes; \email{jfaherty17@gmail.com} 
\and
Department of Astrophysics, 
American Museum of Natural History, Central Park West at 79th Street, New York, NY 10034; \email{jfaherty@amnh.org} 
\and
Department of Engineering Science \& Physics, College of Staten Island, 2800 Victory Blvd., Staten Island, NY 10301 USA
\and
Department of Physics \& Astronomy, Hunter College, 695 Park Avenue, New York, NY 10065, USA
}

\authorrunning{Faherty }

\titlerunning{Young Brown Dwarfs}

\abstract{
Young brown dwarfs and directly-imaged exoplanets have enticingly similar photometric and spectroscopic characteristics, indicating that their cool, low gravity atmospheres should be studied in concert.  Similarities between the peculiar shaped $H$ band, near and mid-IR photometry as well as location on color magnitude diagrams provide important clues about how to extract physical properties of planets from current brown dwarf observations.  In this proceeding we discuss systems newly assigned to 10-150 Myr nearby moving groups, highlight the diversity of this uniform age-calibrated brown dwarf sample, and reflect on their implication for understanding current and future planetary data.
}
\maketitle{}

\section{INTRODUCTION}

Despite different formation mechanisms, brown dwarfs and giant exoplanets share many physical properties, including overlapping temperature regimes and condensate clouds in their atmospheres. Recent studies have revealed a striking resemblance between observations of directly imaged giant exoplanets and young low-temperature brown dwarfs (e.g. \citealt{Faherty13}).  Both populations deviate significantly from older, equivalent temperature objects, and it has been proposed that thick clouds present in the young objects but not in the old ones could explain anomalous observables (\citealt{Barman11}, \citealt{Currie11}, \citealt{Madhusudhan11}). While only a handful of planetary systems can be directly studied with current technology, young brown dwarfs are relatively numerous, bright, and isolated in the field.   They were largely discovered serendipitously while conducting an all-sky or proper motion search for nearby brown dwarfs (e.g. \citealt{Kirkpatrick10}, \citealt{Cruz09}, \citealt{Gizis12}, \citealt{Thompson13}). The current collection lends itself to low, medium and high resolution optical and/or NIR spectroscopy, parallax programs, as well as precise photometric follow-ups. 
As such, they are excellent candidates for extensive studies not currently possible for exoplanets (\citealt{Cruz07, Cruz09}, \citealt{Rice10, Rice10a}, \citealt{Faherty12, Faherty13}, \citealt{Allers13}). In this proceeding we review the characteristics of the low-gravity brown dwarf population and reveal that a number of these sources are indeed members of 10-150 Myr nearby moving groups.

 \section{Evidence for youth}
As discussed in \citealt{Faherty13}, there are several signatures of youth for isolated brown dwarfs.   These can be split into four categories of diagnostic features: photometric, spectroscopic, luminosity, and kinematics.  In the following sub-sections we discuss the broad diagnostics of low-surface gravity brown dwarfs that led to our suspicion that they may belong to nearby young moving groups.

\begin{figure}[h!]
\resizebox{1.0\hsize}{!}{\includegraphics[clip=true]{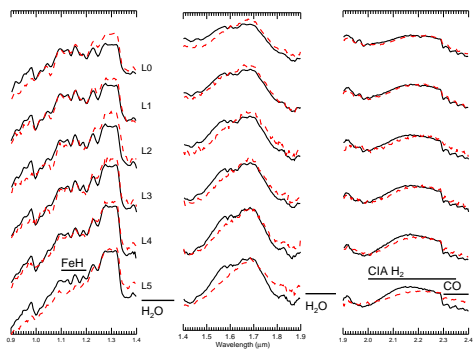}}
\centering
\caption{\scriptsize{
The low resolution NIR L dwarf sequence for normal (black) and $\gamma$ low-gravity (dashed) sources with prominent molecular features highlighted.  At each subtype, a template average of multiple objects sharing the same optical spectral type is shown.  Details on how templates were created will be described in Cruz et al. in prep. and the collection will be available on the BDNYC brown dwarf compendium website (see URL below).}
}
\label{NIRseq}
\end{figure}

\subsection{Spectral Features}
Late-type M and L dwarfs are primarily classified in the optical where spectral features are thought to be most sensitive to temperature variations.  While there are subtle changes indicative of a lower surface gravity in the optical, in order to fully characterize a young source, one must turn to the near-infrared (NIR) where the majority of the flux is emitted and the spectral energy distribution (SED) is more sensitive to gravity variations.  The most telling spectral features of a young isolated brown dwarf are (1) weak, narrow alkali lines, (2) enhanced metal oxide absorption bands and (3) a triangular peaked $H$-band. All of these features are impacted because there is less pressure broadening due to the source's lower surface gravity (see \citealt{Rice11}, \citealt{Allers13} for details).   Consequently, low and medium resolution data in JHK are critical for gauging the youth of an isolated source.  Cruz et al. in prep have created low-resolution NIR spectral templates for field L dwarfs as well as $\gamma$ and $\beta$ (low and intermediate gravity respectively) classifications\footnote{We will be making the low resolution field and low-gravity templates available at the young brown dwarf compendium website http://www.bdnyc.org/young\_bds/} .  Figure 1 shows the L dwarf sequence templates for both normal and $\gamma$ isolated, nearby brown dwarfs.  The triangular shape of the $H$ band is distinct among the low-gravity sources and is prominent (with varying levels of ``extreme") among the 30 sources assigned to moving groups in this proceeding . 

\begin{figure*}[ht!]
\resizebox{0.5\hsize}{!}{\includegraphics[clip=true]{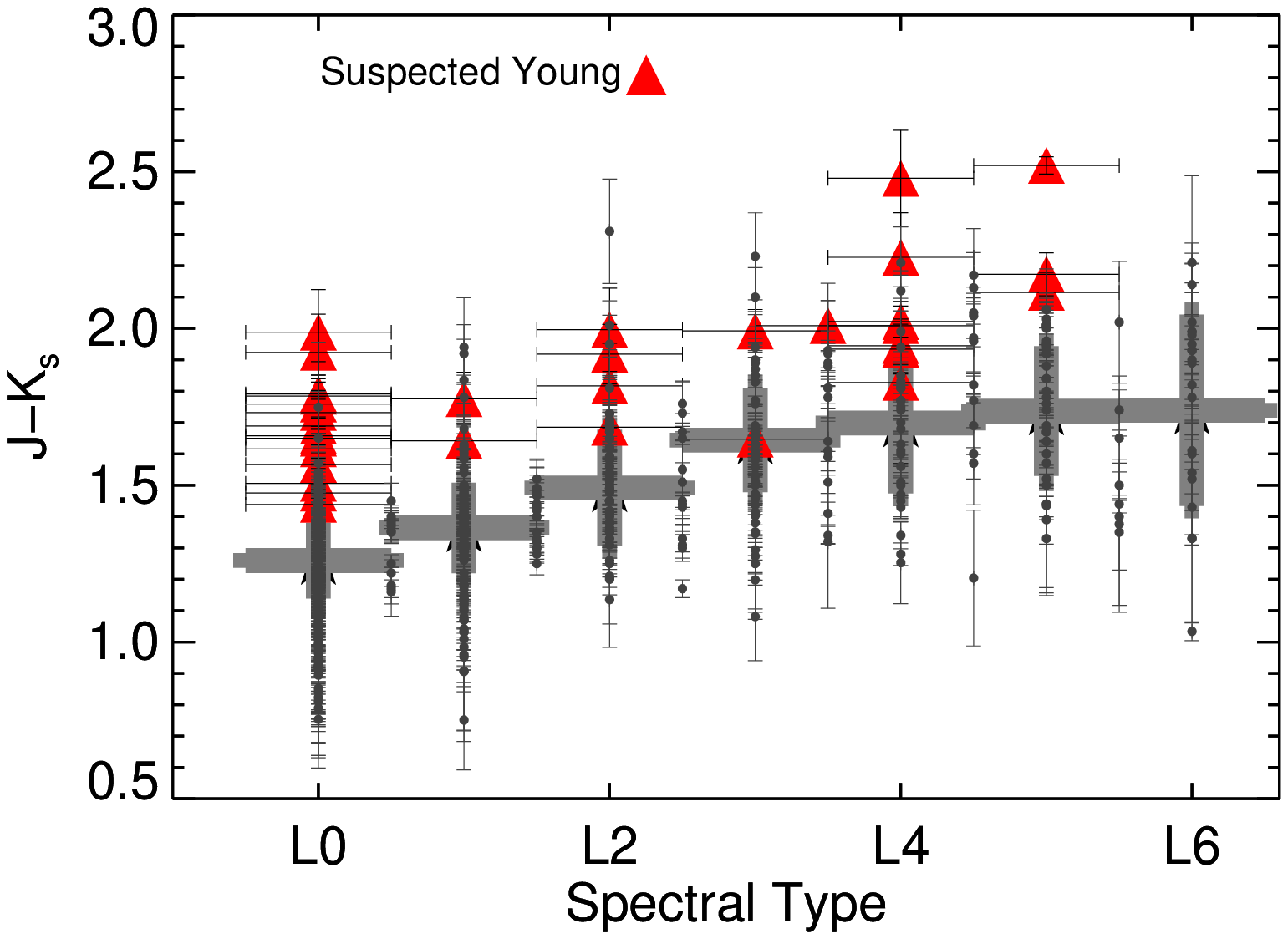}}
\resizebox{0.5\hsize}{!}{\includegraphics[clip=true]{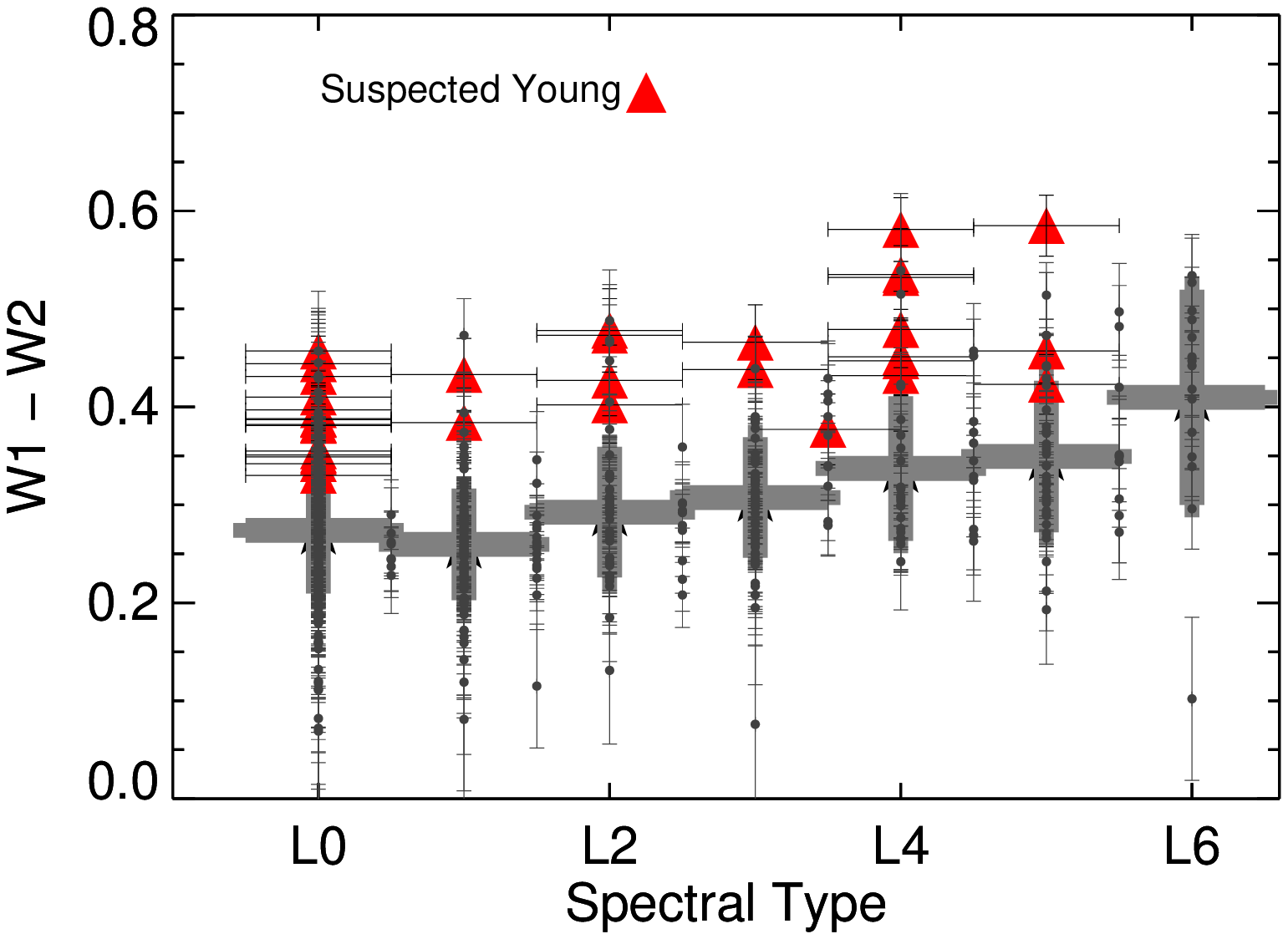}}
\caption{\scriptsize{The NIR (left) and MIR (right) color sequence with the average color for field L dwarfs and their 1$\sigma$ deviation highlighted by a black point and grey box respectively (from \citealt{Faherty13}).  Suspected young $\beta$ and $\gamma$ sources are shown as triangles.  }
 } 
\label{fig:luminosity}
\end{figure*}

\subsection{Photometric Features}
Infrared colors (specifically J-K$_{s}$ and W1-W2) generally increase with decreasing T$_{eff}$ from late-type M through late-type L dwarfs.  Figure 2 shows the overall trends for L dwarfs and demonstrates there is significant dispersion (up to $\sim$1 mag for mid-L dwarfs) among field sources thought to be $>>$ 1 Gyr (\citealt{Faherty09}).  This indicates that variations in secondary parameters---e.g. gravity, metallicity, binarity, atmosphere properties--drive significant diversity throughout the evolution of brown dwarfs.  Isolating only sources designated as $\gamma$ or $\beta$ gravity (see Figure 2), one finds that this population deviates from field equivalents by more than 2$\sigma$.  \citet{Faherty13} find that $\gamma$ sources are up to 0.8 mag redder in (J-K$_{s}$) and 0.25 mag redder in (W1-W2).   Indeed, to date,  there is no known source classified as low-surface gravity with a NIR or mid-infrared (MIR) color that is blueward of the median for an equivalent subtype.  Physically, the red NIR color in young objects is brought on by enhanced photospheric dust as well as gravity-induced changes to broadband NIR features.  

\subsection{Luminosity Features}
Parallax programs in recent years have begun to fill in the gap of our understanding of luminosity features for young isolated brown dwarfs (\citealt{Faherty12}, \citealt{Liu13}).  Emerging as a perplexing trend is that while young M dwarfs appear overluminous as expected on color-magnitude diagrams, suspected young L dwarfs are  underluminous in the NIR for their assigned spectral type (Figure 3).  At least two factors could contribute to this trend: 1) Spectral types assigned to low gravity objects do not necessarily correspond (in temperature or luminosity) to spectral types assigned for field dwarfs. Low gravity significantly affects the entire spectrum so field and young L dwarfs with the same spectral type have quite different spectra (e.g., an L1 and L1$\gamma$, see \citealt{Cruz09}). As a result, it is quite likely that the low gravity L dwarfs have a cooler spectral type/temperature relation, thus making them appear underluminous on a spectral-type/absolute magnitude diagram  2) Young objects are dustier than field-aged dwarfs and thicker clouds shift flux to longer wavelengths (e.g. 2M0355--see \citealt{Faherty13}).

\subsection{Tangential Velocities}
Individual space motions cannot be used to date an object.  However, one can look to a population for statistical deviations that can be used as an age indicator.  In \citealt{Faherty09}, both red and blue photometric outliers were isolated, and their kinematics examined as subpopulations against the field.   Redder objects demonstrated smaller mean velocities and tighter dispersions, while blue objects demonstrated the converse.   Isolating the low-surface gravity sources, that happen to be among the reddest in the brown dwarf sample, kinematics show [v$_{tan}$, $\sigma_{tan}$]=[18 , 15] km s$^{-1}$, the smallest numbers for a subpopulation of the full kinematic analysis.  Using a crude age-velocity analysis indicates that the low-gravity population is indeed younger than the mean population.
 \begin{figure}[]
\resizebox{1.0\hsize}{!}{\includegraphics[clip=true]{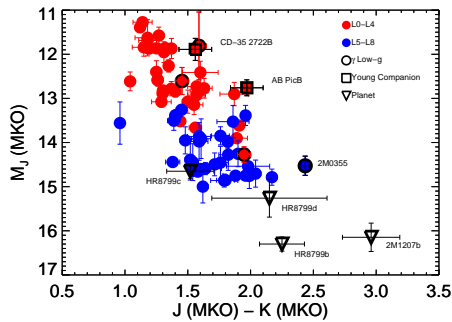}}
\caption{\scriptsize{
The NIR color magnitude diagram for brown dwarfs supplemented with planetary mass companions, young low-mass companions, and $\gamma$ L dwarfs. }
}
\label{NIRseq}
\end{figure}

\section{Assigning Ages to Low-Gravity Brown Dwarfs}
We have identified 65 low-surface gravity brown dwarfs (e.g. those with features deviant in 3-4 of the categories described above) that are candidates for an age calibrated sample.  As shown in Figure 4, many are coincident with the locations of the Argus ($\sim$ 30 Myr),  $\beta$ Pictoris ($\sim$10 Myr), TW Hydrae ($\sim$10 Myr), Tucana Horologium ($\sim$30 Myr), or AB Doradus ($\sim$ 150 Myr) associations. To conclusively assign membership, we require precise astrometric measurements.  Consequently, as part of the Brown Dwarf Kinematics Project (BDKP--\citealt{Faherty09, Faherty10, Faherty11, Faherty12}), we have been collecting parallax, proper motion and radial velocity data on the full brown dwarf population.  We have prioritized the low-surface gravity sources and recently published precise parallaxes and proper motions for a subset (\citealt{Faherty09, Faherty12, Faherty13}) and expect to publish radial velocities in a forthcoming paper (Faherty et al. in prep).  In total we have enough kinematic information on our sources to compute (or make estimates of) space velocity and positions to assign preliminary membership to nearby moving groups.     
 \section{Diversity of Young Brown Dwarfs}
Using estimated $UVW$ velocities and $XYZ$ positions in combination with a convergent point and Bayesian analysis (Rodriguez et al. 2013; \citealt{Malo13}) we assign membership to 30 low-surface gravity brown dwarfs.  We list the groups for which we are assigning membership in Table 1 along with the optical gravity classification ($\gamma$ for low and $\beta$ for intermediate) for the sources.  As shown, we see a diversity among the gravity classifications assigned to groups with a uniform age. For instance, $\beta$ Pictoris, a $\sim$ 10 Myr old moving group has eight $\gamma$ and four $\beta$ sources.  \citet{Allers13} discuss this issue in context with the $\sim$150 Myr AB Doradus members 2M0355 and CD35-2722B which look very different in the NIR--the former a $\gamma$ and the latter a $\beta$.  The diversity that is emerging among equal age sources is not surprising.  \citet{Faherty13} discuss the $\gamma$ L dwarfs and find that these proposed lowest surface gravity isolated brown dwarfs show a large range in the extent of their red photometric color (in both the NIR and the MIR).  Indeed in Figure 2, the $\gamma$ and $\beta$ gravity sources show almost a 1 magnitude range in outlier color in the NIR and 0.2 magnitude range in outlier color in the MIR for equivalent subtypes.   For those sources with parallax measurements we find that the M dwarfs are overluminous in the NIR whereas the L dwarfs are normal to underluminous in the NIR regardless of the age calibration.   We postulate that this diversity among the age calibrated sample is due to complex atmospheric chemistry which can differ from source to source.

\section{Discussion:  Connection to Exoplanets}
The estimated T$_{eff}$ of the directly imaged planetary-mass companions 2M1207b ($\sim$ 10 Myr) and HR8799b ($\sim$ 30 Myr) are $\sim$1100K and 1600K, respectively, corresponding to mid L and early T spectral types (\citealt{Barman11}, \citealt{Skemer11}). Detailed studies of their near-IR spectra and photometric data reveal that both planets (1) are 1-2 mag underluminous on color-magnitude diagrams, (2) have unusually red near-IR colors, and (3) display sharply peaked H-band spectra.   Our population of 30 brown dwarfs, now kinematically linked to  10-150 Myr associations share in these deviations from field brown dwarfs.  

Emerging as the most probable explanation for the exoplanet observables are enhanced clouds induced by gravity effects.  Independently, the same explanation has been reached for the young brown dwarfs (\citealt{Faherty13}).  Hence atmosphere and age induced features are tangled in this low-temperature regime.  Based on the diversity of the young brown dwarf sample discussed here-in, there is much work to be done in deciphering what features can be attributed strictly to gravity and which to clouds.   Fortunately there is detailed information on the young brown dwarfs.  Future work will include a careful and systematic study of the large amount of data in hand and a thorough comparison to existing atmosphere and evolution models in order to tease out the physics of deviant observables and inform the next generation of models.

 \begin{figure}[t!]
\resizebox{1.0\hsize}{!}{\includegraphics[clip=true]{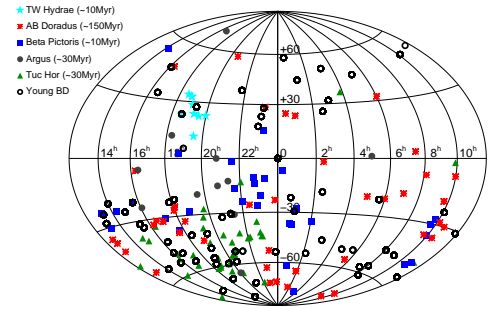}}
\caption{\scriptsize
The positions of young stars in nearby moving groups as well as suspected young brown dwarfs in an all-sky galactic coordinate aitoff projection (Faherty et al. in prep).
}
\label{eta}
\end{figure}

\begin{table}
\caption{\scriptsize{Brown Dwarfs in Nearby Moving Groups}}
\label{abun}
\begin{tabular}{lccccccc}
\hline
\\
Group & Age & $\#$ BD & $\gamma$ & $\beta$  \\
\hline
\\
Argus  &$\sim$30 Myr &2&0&2\\

AB Doradus  &$\sim$150 Myr&  4&2&2   \\

$\beta$ Pictoris  &$\sim$ 10 Myr &  12&8&4\\

Tucana Horlogium  &$\sim$ 30 Myr &  12&10&2\\
\\
\hline
\end{tabular}
\caption{\scriptsize
We list the number of  $\gamma$ and $\beta$ gravity brown dwarfs (low and intermediate respectively) that we have assigned to nearby moving groups.  Thirty in total have been assigned to 10-150 Myr groups.  
}

\end{table}

\bibliographystyle{aa}
\bibliography{paper2}

\begin{thebibliography}{20}
\expandafter\ifx\csname natexlab\endcsname\relax\def\natexlab#1{#1}\fi

\bibitem[{{Allers} \& {Liu}(2013)}]{Allers13}
{Allers}, K.~N. \& {Liu}, M.~C. 2013, ArXiv e-prints

\bibitem[{{Barman} {et~al.}(2011){Barman}, {Macintosh}, {Konopacky}, \&
  {Marois}}]{Barman11}
{Barman}, T.~S., {Macintosh}, B., {Konopacky}, Q.~M., \& {Marois}, C. 2011,
  \apj, 733, 65

\bibitem[{{Cruz} {et~al.}(2009){Cruz}, {Kirkpatrick}, \& {Burgasser}}]{Cruz09}
{Cruz}, K.~L., {Kirkpatrick}, J.~D., \& {Burgasser}, A.~J. 2009, \aj, 137, 3345

\bibitem[{{Cruz} {et~al.}(2007){Cruz}, {Reid}, {Kirkpatrick}, {Burgasser},
  {Liebert}, {Solomon}, {Schmidt}, {Allen}, {Hawley}, \& {Covey}}]{Cruz07}
{Cruz}, K.~L., {Reid}, I.~N., {Kirkpatrick}, J.~D., {et~al.} 2007, \aj, 133,
  439

\bibitem[{{Currie} {et~al.}(2011){Currie}, {Burrows}, {Itoh}, {Matsumura},
  {Fukagawa}, {Apai}, {Madhusudhan}, {Hinz}, {Rodigas}, {Kasper}, {Pyo}, \&
  {Ogino}}]{Currie11}
{Currie}, T., {Burrows}, A., {Itoh}, Y., {et~al.} 2011, \apj, 729, 128

\bibitem[{{Faherty} {et~al.}(2011){Faherty}, {Burgasser}, {Bochanski},
  {Looper}, {West}, \& {van der Bliek}}]{Faherty11}
{Faherty}, J.~K., {Burgasser}, A.~J., {Bochanski}, J.~J., {et~al.} 2011, \aj,
  141, 71

\bibitem[{{Faherty} {et~al.}(2009){Faherty}, {Burgasser}, {Cruz}, {Shara},
  {Walter}, \& {Gelino}}]{Faherty09}
{Faherty}, J.~K., {Burgasser}, A.~J., {Cruz}, K.~L., {et~al.} 2009, \aj, 137, 1

\bibitem[{{Faherty} {et~al.}(2012){Faherty}, {Burgasser}, {Walter}, {Van der
  Bliek}, {Shara}, {Cruz}, {West}, {Vrba}, \& {Anglada-Escud{\'e}}}]{Faherty12}
{Faherty}, J.~K., {Burgasser}, A.~J., {Walter}, F.~M., {et~al.} 2012, \apj,
  752, 56

\bibitem[{{Faherty} {et~al.}(2010){Faherty}, {Burgasser}, {West}, {Bochanski},
  {Cruz}, {Shara}, \& {Walter}}]{Faherty10}
{Faherty}, J.~K., {Burgasser}, A.~J., {West}, A.~A., {et~al.} 2010, \aj, 139,
  176

\bibitem[{{Faherty} {et~al.}(2013){Faherty}, {Rice}, {Cruz}, {Mamajek}, \&
  {N{\'u}{\~n}ez}}]{Faherty13}
{Faherty}, J.~K., {Rice}, E.~L., {Cruz}, K.~L., {Mamajek}, E.~E., \&
  {N{\'u}{\~n}ez}, A. 2013, \aj, 145, 2

\bibitem[{Gizis {et~al.}(2012)Gizis, Faherty, Liu, Castro, Shaw, Vrba, Harris,
  Aller, \& Deacon}]{Gizis12}
Gizis, J.~E., Faherty, J.~K., Liu, M.~C., {et~al.} 2012, arXiv.org

\bibitem[{{Kirkpatrick} {et~al.}(2010){Kirkpatrick}, {Looper}, {Burgasser},
  {Schurr}, {Cutri}, {Cushing}, {Cruz}, {Sweet}, {Knapp}, {Barman},
  {Bochanski}, {Roellig}, {McLean}, {McGovern}, \& {Rice}}]{Kirkpatrick10}
{Kirkpatrick}, J.~D., {Looper}, D.~L., {Burgasser}, A.~J., {et~al.} 2010,
  \apjs, 190, 100

\bibitem[{{Liu} {et~al.}(2013){Liu}, {Dupuy}, \& {Allers}}]{Liu13}
{Liu}, M.~C., {Dupuy}, T.~J., \& {Allers}, K.~N. 2013, Astronomische
  Nachrichten, 334, 85

\bibitem[{{Madhusudhan} {et~al.}(2011){Madhusudhan}, {Burrows}, \&
  {Currie}}]{Madhusudhan11}
{Madhusudhan}, N., {Burrows}, A., \& {Currie}, T. 2011, \apj, 737, 34

\bibitem[{{Malo} {et~al.}(2013){Malo}, {Doyon}, {Lafreni{\`e}re}, {Artigau},
  {Gagn{\'e}}, {Baron}, \& {Riedel}}]{Malo13}
{Malo}, L., {Doyon}, R., {Lafreni{\`e}re}, D., {et~al.} 2013, \apj, 762, 88

\bibitem[{{Rice} {et~al.}(2010{\natexlab{a}}){Rice}, {Barman}, {Mclean},
  {Prato}, \& {Kirkpatrick}}]{Rice10a}
{Rice}, E.~L., {Barman}, T., {Mclean}, I.~S., {Prato}, L., \& {Kirkpatrick},
  J.~D. 2010{\natexlab{a}}, \apjs, 186, 63

\bibitem[{{Rice} {et~al.}(2011){Rice}, {Faherty}, {Cruz}, {Barman}, {Looper},
  {Malo}, {Mamajek}, {Metchev}, \& {Shkolnik}}]{Rice11}
{Rice}, E.~L., {Faherty}, J.~K., {Cruz}, K., {et~al.} 2011, ArXiv e-prints

\bibitem[{{Rice} {et~al.}(2010{\natexlab{b}}){Rice}, {Faherty}, \&
  {Cruz}}]{Rice10}
{Rice}, E.~L., {Faherty}, J.~K., \& {Cruz}, K.~L. 2010{\natexlab{b}}, \apjl,
  715, L165

\bibitem[{{Skemer} {et~al.}(2011){Skemer}, {Close}, {Sz{\H u}cs}, {Apai},
  {Pascucci}, \& {Biller}}]{Skemer11}
{Skemer}, A.~J., {Close}, L.~M., {Sz{\H u}cs}, L., {et~al.} 2011, \apj, 732,
  107

\bibitem[{{Thompson} {et~al.}(2013){Thompson}, {Kirkpatrick}, {Mace},
  {Cushing}, {Gelino}, {Griffith}, {Skrutskie}, {Eisenhardt}, {Wright},
  {Marsh}, {Mix}, {Beichman}, {Faherty}, {Toloza}, {Ferrara}, {Apodaca},
  {McLean}, \& {Bloom}}]{Thompson13}
{Thompson}, M.~A., {Kirkpatrick}, J.~D., {Mace}, G.~N., {et~al.} 2013, ArXiv
  e-prints

\end{thebibliography}

\end{document}